\documentclass[12pt,pra,reprint,superscriptaddress,amsmath,amsfonts,amssymb,nofootinbib]{revtex4-1}

\usepackage[utf8]{inputenc}
\usepackage[T1]{fontenc}
\usepackage[sc,osf]{mathpazo}
\usepackage{amsmath, amsthm, amssymb,amsfonts,mathbbol,amstext}
\usepackage{graphicx}
\usepackage{dcolumn}
\usepackage{bm}
\usepackage{bbm}
\usepackage{mathtools}
\usepackage{comment}
\usepackage{multirow}
\usepackage{diagbox}
\usepackage{float}

\usepackage[pretty]{revquantum}
\usepackage[english]{babel}
\usepackage{graphicx}
\usepackage[caption=false]{subfig}
\usepackage{mathrsfs}
\usepackage{amsmath}
\usepackage{amsthm}
\usepackage{dsfont}
\usepackage{thmtools, thm-restate}
\usepackage{bm}
\usepackage{xcolor}
\usepackage{hyperref}
\usepackage{hypernat}
\usepackage[utf8]{inputenc}
\usepackage[T1]{fontenc}
\usepackage{cleveref}
\usepackage{soul}

\declaretheorem[parent=section,name=Theorem]{thm}

\declaretheorem[style=definition,sibling=thm]{definition}

\declaretheorem[sibling=thm]{proposition}

\newcommand{\channel}[2]{\mathcal{N}_{#1|#2}}

\newcommand{\state}[2]{\varrho_{#1|#2}}
\newcommand{\ketbra}[2]{\ket{#1}\bra{#2}}

\usepackage{xcolor}

\newcommand{\soplinear}[1]{\mathcal{L}(\mathcal{H}_{#1})} 

\newcommand{\kb}[2]{\ket{#1}\bra{#2}} 
\newcommand{\kbt}[4]{\ket{#1}\bra{#2} \otimes \ket{#3}\bra{#4}} 
\newcommand{\rc}[2]{\rho_{#1|#2}} 
\newcommand{\vrc}[2]{\varrho_{#1|#2}} 
\newcommand{\kbtre}[4]{Tr(E_{y}^{B}\ket{#3}\bra{#4})\ket{#1}\bra{#2}} 
\newcommand{\rcbayes}[4]{\rho_{#1 | #2} \star (\rho_{#3} \rho_{#4}^{-1})} 

\newcommand{\rcc}[2]{\varrho_{#1|#2}}

\newcommand{\patrace}[1]{\text{Tr}(#1)}

\newcommand{\ttensor}[2]{#1\otimes #2}

\begin{document}

\title{Recasting Schrödinger's Cat Thought Experiment as a Remote Measurement Problem}

\author{Lucas L. Brugger}
\email{lucasbrugger7@gmail.com}
\affiliation{Universidade Federal de Juiz de Fora, Departamento de Física, Juiz de Fora, MG, Brasil.}
\author{Cristhiano Duarte}
\email{cristhianoduarte@gmail.com}
\affiliation{Universidade Federal de Juiz de Fora, Departamento de Física, Juiz de Fora, MG, Brasil.}
\affiliation{Institute for Quantum Studies, Chapman University, One 
University Drive, Orange, CA, 92866, USA}
\affiliation{Instituto de Física, Universidade Federal da Bahia, Campus de Ondina, Rua Barão do Geremoabo, s.n., Ondina, Salvador, BA 40210-340, Brazil}%
\affiliation{Fundação Maurício Grabois, R. Rego Freitas, 192 - República, São Paulo - SP, 01220-010, Brazil}
\author{Bruno F. Rizzuti}
\email{brunorizzuti@ufjf.br}
\affiliation{Universidade Federal de Juiz de Fora, Departamento de Física, Juiz de Fora, MG, Brasil.}

\date{\today}

\begin{abstract}

With 2025 being declared the Year of Quantum Science and Technology, our contribution seeks to provide a fresh perspective on Schrödinger's cat thought experiment. We reinterpret this experiment by viewing it through the lens of quantum theory as a generalisation of classical probability, rooted in a Bayesian subjectivist framework. In this revised approach, we treat the experiment as a remote measurement problem. Specifically, we explore how the beliefs of two agents, Alice and Bob, who are spatially separated yet share a quantum state, are updated when local measurements are conducted on their respective systems. Through this reinterpretation of the well-known experiment, we also aim to offer an educational perspective that will be beneficial for young scientists interested in the field of quantum theory.

\end{abstract}
\maketitle
\section{Introduction} \label{Sec.Intro}

Contemporary pop culture is marred with direct and indirect references to Schrödinger's cat. It does not matter the medium, be it a book, a newspaper, entertainment media in any form, your favourite science communicator online, or a mere conversation among friends; there is a fair chance that a possibly alive, possibly dead cat inside a box might have already been mentioned a couple of times. The popularity of Schrödinger’s Cat, the thought experiment proposed by Erwin Schrödinger in 1935 \cite{trimmer1980present}, undeniably stems from the intriguing features of a new theory that emerged at the beginning of the 20th century~\cite{nielsen_chuang_2010,cohen}. Quantum Theory, this new physical theory, named after Max Planck's and Albert Einstein's seminal works \cite{cohen,nielsen_chuang_2010}, was developed to explain experimental results that classical mechanics could not properly account for~\cite{cohen}.

In the following years, the development of quantum theory accelerated, with many distinguished researchers making significant contributions to the understanding of puzzling new phenomena: entanglement \cite{einstein1935can,trimmer1980present}, no-cloning \cite{WoottersZurek1982,Dieks1982}, no-broadcasting \cite{BarnumEtAl1996}, Bell non-locality \cite{Bell1964,Bell66} and steering \cite{UolaCostaNguyenGuehne2020}, just to name a few. Over time, the theory has extended far beyond academic debate or science fiction, demonstrating an incredible ability to be applied to the engineering of an ever-increasing list of gadgets---from quantum computers themselves to quantum networks for communication and medical equipment~\cite{Roadmap_Quantum_Nanotechnologies_2020,aslam2023quantum,Watts2021PhotonQuantumEntanglement}.

The foundational work advanced by Werner Heisenberg \cite{VanDerWaerden1967}, Max Born and Pascual Jordan \cite{VanDerWaerden1967} and Erwin Schrödinger \cite{Schrodinger1928Collected},\footnote{It is important to emphasise that Schrödinger's seminal works that shed light on wave mechanics were developed through 1925 but only published in 1926.} together with the profound impact of quantum theory on modern life, motivated UNESCO to declare 2025 as the International Year of Quantum Science and Technology \cite{jornal_unesp}. One might argue that quantum theory does not have a single, fixed `birth certificate'; nonetheless, it is undeniable that 2025 marks the centenary of several pivotal breakthroughs in its development.

In any case, since its inception, the defining feature of quantum theory has been its intrinsic probabilistic nature \cite{LS13,leifer2006quantum,Fano1957,caves2002quantum,barnett2000bayes}, a central point of intense debate to this day \cite{Cabello2017}. Inspired by this distinctive character, numerous interpretations of quantum theory have emerged over the decades. From realist to anti-realist, several strikingly different standpoints co-inhabit the realm of the foundations of quantum mechanics \cite{Cabello2017}. For a variety of reasons, which we will not have time to address in any detail in this work, the Copenhagen interpretation~\cite{Cabello2017,cohen} itself and its inspired re-interpretations, now lumped together under the term Copenhagen-ish, are getting traction and attracting some attention in the community of foundations of quantum mechanics \cite{SYL25}. 

Writing agency in the heart of quantum theory, Copenhagen-ish interpretations house those standpoints that deny that the wavefunction is an intrinsic, actual property of an individual quantum system. We could say even more: Copenhagen-ish interpretations tend to avoid attributing intrinsic ontological
properties to physical systems described by quantum theory \cite{SYL25}. Eschewing away solipsism \cite{Cavalcanti21}, this strand of interpretations does not deny the existence of an objective world outside the consciousness of observers, however, according to it, quantum theory does not deal directly with intrinsic properties of the observed system, but with the experiences an observer (an agent) has of the observed system. This type of interpretation can be epistemic, primarily, in two senses: they are about knowledge if they view the quantum state as an observer’s knowledge on the results of future experiments, or about belief\footnote{Belief, as referred to in the text, can be characterized as a quantification that some agent analyzing a certain system attributes to the probability of an event occurring \cite{HantiLinbayesian2022}.} if they view the quantum state as an agent’s expectations about the results of future actions. In either case, some radical strands advanced within the umbrella Copenhagen-ish interpretation want to draw a parallel with subjective standpoints of classical probability~\cite{Ramsey31,Finetti78, Fishburn1986}. 

It is precisely this parallel we want to push further in this work. Under the assumption that quantum theory, viewed as a theory of probabilistic assignments, can be understood as a generalisation of classical probability, we will explore Schrödinger's thought experiment using the mathematical framework of Quantum Conditional States \cite{leifer2006quantum, LS13}---a framework devised to cast quantum theory as a potentially neutral theory of Bayesian inference. In doing so, we will reassess Schrödinger’s famous cat and reframe it under the lens of a remote measurement. A due homage to a widely known and, to this day, thought-provoking scenario.

For this reason, this work is organised as follows. In Section~\ref{Sec.Formalism}, we begin by introducing the formalism that will be used to develop our recasting of Schrödinger’s cat thought experiment. Section~\ref{SubSec.Originalexperiment} provides a brief explanation of the original thought experiment, followed by the presentation of our proposed recasting. The corresponding mathematical development is then detailed in Subsection~\ref{SubSec.mathdev}. Finally, Section~\ref{Sec.Conclusion} presents our conclusions.

\section{Conditional States Approach}\label{Sec.Formalism}


Although versions of the conditional states approach (CSA) had definitively appeared in the literature before \cite{LP08,Leifer06}, its mature form, the one we chose to adopt in this work, was first fleshed out in detail by the authors of \cite{LS13} and \cite{LS14}. In those works, in an attempt to create a formal parallel with classical Bayesian inference, the authors advanced the stance that quantum theory may be, to some extent, viewed as a neutral theory of Bayesian inference. Under the CSA, several results in classical decision theory and causal inference naturally find their counterpart in the quantum realm \cite{LS14, LD22}---often making explicit the dependence of those results on the underlying set-theoretical structure on top of which we build (generalised) probabilistic theories \cite{LD22}.

Beyond drawing parallels between classical and quantum inference problems, the CSA is also useful within quantum theory itself. As we will argue in this contribution, through the conditional states language, it is possible to recast some thought-provoking quantum scenarios into situations that we have a stronger understanding of, not only shedding new light on older phenomena but also making them potentially clearer. With that being said, it is crucial to emphasise that the program initiated in \cite{LS13} and \cite{LS14} as a whole suffers from some unavoidable subtleties for multipartite systems---as noticed, not least, by the very authors of those papers. Because we consider a scenario involving only two agents, those problems do not affect our work.

\subsection{A Glimpse of the Conditional States Approach}

To provide a glimpse of the CSA, we must first introduce a fundamental concept that underlies the entire approach: the concept of a \textit{region}. Broadly speaking, a \textit{region} is a portion of space-time where an agent can perform a single intervention on a system. To such a portion, say region $A$, we associate a Hilbert space $\mathcal{H}_{\text{A}}$. In stark contrast to the usual textbook approach, in the CSA the Hilbert space arising from an evolution through a CPTP channel\footnote{In this work, $\mathcal{L}(\mathcal{H}_{X})$ denotes the collection of all linear operators over the Hilbert space $\mathcal{H}_{X}$.}, $\mathcal{N}:\mathcal{L}(\mathcal{H}_{\text{in}}) \rightarrow \mathcal{L}(\mathcal{H}_{\text{out}})$, is modeled as a \textit{composed region}, a tensor product \cite{WatrousLectureNotes,nielsen_chuang_2010} of the initial system's and the future\footnote{The term `future' is used here not only in the temporal sense, but also in the sense of a causal future, that is, one in which one region is brought to another via a CPTP channel.} system's Hilbert spaces, that is $\mathcal{H}_{in} \otimes \mathcal{H}_{out}$. As usual, $\mathcal{L}(\mathcal{H})$ denotes the space of linear operators acting on the Hilbert space $\mathcal{H}$. By doing this move, usual composite quantum systems are treated on the same footing in CSA: each system an agent can interact with in a multipartite quantum system determines a quantum region, and the composed region is the tensor product of each individual Hilbert space. In particular, the Hilbert space associated with a bipartite system, with a bipartite region, is $\mathcal{H}_{A} \otimes \mathcal{H}_{B}$, where $\mathcal{H}_{A}$ ($\mathcal{H}_{B}$) is the Hilbert space associated with a region $A$ ($B$).
%

To treat both causal and acausal scenarios briefly discussed above within the CSA framework, it is necessary to introduce a paradigmatic isomorphism between physical processes and states. In our work, it assumes the following form~\cite {Jamiolkowski72}.

\begin{definition}[Jamio\l kowski Isomorphism]

Let 
\begin{equation}
\mathcal{N}:\mathcal{L}(\mathcal{H}_{A}) \rightarrow \mathcal{L}(\mathcal{H}_{B})
\end{equation}
be a linear map where $\mathcal{H}_{A}$ and $\mathcal{H}_{B}$ are finite-dimensional Hilbert spaces. 
The \emph{(Choi-)Jamio\l kowski image} of $\mathcal{N}$ is the operator $\rho \in \mathcal{L}(\mathcal{H}_A \otimes \mathcal{H}_B)$ defined as
    \begin{equation}
        \rho := (\mbox{id}_{A} \otimes \mathcal{N} \circ \mbox{T}_{A}) \sum_{i,j = 1}^{d_A}\ket{i}\bra{j}_{A} \otimes\ket{i}\bra{j}_{A},
        \label{Eq.DefCJIsomorphism}
    \end{equation}
where $d_A=\mbox{dim}(\mathcal{H}_{A})$ and the transposition $\mbox{T}_{A}$ is taken with respect to some basis in $\mathcal{H}_{A}$. The action of $\mathcal{N}$ on $\mathcal{L}(\mathcal{H}_{A})$ is given by
    \begin{equation}
        \mathcal{N}(\sigma_{A})=\mbox{Tr}_{A}[\rho ( \sigma_{A} \otimes \mathds{1}_{B})],
    \end{equation}
where $\mathcal{N}(\sigma_{A}) \in \mathcal{L}(\mathcal{H}_{B})$. 
\label{Def.JamilIso}
\end{definition}
The fact that $\rho$ and $\mathcal{N}$ are isomorphically connected can be directly verified, since
\begin{align}\nonumber
    &\mbox{Tr}_{A}[\rho(\sigma_{A} \otimes \mathds{1}_{B})] = \nonumber \\ \nonumber
    &=\mbox{Tr}_{A}\left[ (\mbox{id} \otimes \mathcal{N})(\sum_{i,j}^{d} \ketbra{i}{j} \otimes \ketbra{j}{i}) (\sigma_{A} \otimes \mathds{1}_{B}) \right] \\ \nonumber
    &= \sum_{i,j}^{d}\mbox{Tr}_{A}\left[ \ketbra{i}{j}\sigma_{A} \otimes \mathcal{N}(\ketbra{j}{i}) \right] = \sum_{i,j} \bra{j}\sigma_{A}\ket{i} \mathcal{N}(\ketbra{j}{i})  \\
    &=\sum_{i,j}^{d} (\sigma_{A})_{j,i}\mathcal{N}(\ketbra{j}{i}) =\mathcal{N}(\sigma_{A}). 
\end{align}

As we will see in a moment, the idea behind the Jamio\l kowski isomorphism, and of the CSA in particular, is that it maps any CPTP map into a (non-normalised) bipartite state.\footnote{There still is a debate in the literature about whether the transposition map should appear in expression~\eqref{Eq.DefCJIsomorphism}---see Refs.~\cite{MKPM19,OCB12,CDAP09}.  Different authors are more inclined towards one or another, but in this work, we will stick to the def.~\ref{Def.JamilIso} above, following Refs.~\cite{LS13, LS14}.} This perspective is formalised in the result below, whose proof follows after the main text, in Appendix \ref{App.ProofIsomorphism}.

\begin{proposition}\label{Prop.conditionalstaterule}
Let $\mathcal{N}:\mathcal{L}(\mathcal{H}_{A}) \rightarrow \mathcal{L}(\mathcal{H}_{B})$ be a linear map, and let $\rho \in \mathcal{L}(\mathcal{H}_{B} \otimes \mathcal{H}_{A})$ be the Jamio\l kowski isomorphic operator associated to it. It follows that $\rho$ satisfies
\begin{enumerate}
    \item[(a)] $\rho^{{T}_{A}} \geq 0$
    \item[(b)] $\mbox{Tr}_{B}[\rho]=\mathds{1}_{A}$
\end{enumerate}
if, and only if, $\mathcal{N} \circ \mbox{T}_A$ is a completely positive and trace preserving map.
\label{Prop.StateChoi}
\end{proposition}

Because we are focusing on physical processes, we will centre our attention on quantum channels rather than arbitrary linear maps. In other words, we want to demand complete positivity from $\mathcal{N}$ rather than from the composite $\mathcal{N} \circ \mbox{T}_{A}$. In so doing, Prop.~\ref{Prop.StateChoi} remains applicable, but it needs a small adaptation. If we start with a CPTP map $\mathcal{C}$ whose Jamio\l kowski-isomorphic image is $\varrho$, by defining $\rho := \varrho^{T_{A}}$, we know that $\mathcal{C} \circ \mbox{T}_{A}$ is its Jamio\l kowski-isomorphic image, therefore (via Prop.~\ref{Prop.StateChoi}) the complete positivity of $\mathcal{C} = \mathcal{C} \circ \mbox{T}_{A} \circ \mbox{T}_{A} $ is equivalent to (a') $\varrho= (\varrho^{T_{A}})^{T_{A}} \geq 0 $ and (b') $\mbox{Tr}_{B}[\varrho]=\mathds{1}_{A}$ This is why we stick with $\varrho$, as it reflects directly the CPTP-ness of a given quantum channel.  

To strengthen the connection between conditional states, physical processes, conditional probability and regions, throughout this work, we will consistently use a suggestive index notation for both states and channels. We will write time steps and Hilbert spaces' labels in a manner reminiscent of the `given' notation commonly used for conditional probabilities. Thus, when we write $\channel{B}{A}$, it is meant to imply, first, that  
    \begin{equation}
        \mathcal{N}_{B|A}:\mathcal{L}(\mathcal{H}_{A}) \rightarrow \mathcal{L}(\mathcal{H}_{B}),
    \end{equation}
and, secondly, that thought as a physical process, it determines the dynamics from time step $t_A$ to time step $t_B$---with $t_B \geq t_{A}$. Analogously, the (Choi-Jamio\l kowski) \emph{conditional state} associated with this channel will be written as $\varrho_{B|A}$, where
    \begin{equation}
        \varrho_{B|A} \in \mathcal{L}(\mathcal{H}_{B} \otimes \mathcal{H}_{A}).
    \end{equation}
This notation not only enhances readability, but it also creates a parallel with other works where expressions like $B|A$ have a particular Bayesian-probabilistic meaning---see Refs.~\cite{LS13,LS14}. For example, by adopting this notation we can promptly observe a direct connection with classical belief propagation, one of the tenets of Bayesian inference. Specifically, given an state of a region $A$, denoted $\rho_{A}$, and a CPTP channel $\varepsilon_{B|A}: \soplinear{A} \mapsto \soplinear{B}$, we have
\begin{equation}\label{Ex.channelprop}
    \rho_{B} = \varepsilon_{B|A}(\rho_A).
\end{equation}
Recalling def.~\ref{Def.JamilIso} the expression \eqref{Ex.channelprop} can be rewritten as,
\begin{equation}\label{Def.beliefpropagation}
    \rho_{B} = \text{Tr}_{A}(\varrho_{B|A}\rho_{A}),
\end{equation}
where expression \eqref{Def.beliefpropagation} is called within the CSA as \textit{quantum belief propagation rule} \cite{LS13}.\footnote{Note that for expression~\eqref{Def.beliefpropagation} to remain consistent, the product inside the trace is of the form $\varrho_{B|A}(\rho_{A} \otimes \mathds{1}_{B})$. Nevertheless, here the identity $\mathds{1}_{B}$ is suppressed to provide a more compact notation.} Also recall that the classical belief propagation is usually given by 
\begin{align}\label{Def.ClassicalBeliefPropagation}
   P(B)=\sum_{a \in Out(A)}P(B|A=a)P(A=a), 
\end{align}
where $P(B|A)$ is the conditional probability of $B$ given $A$. Although apparently naive, it is the similarity between Eq.~ \eqref{Def.beliefpropagation} and Eq.~ \eqref{Def.ClassicalBeliefPropagation} we use to justify dubbing $\varrho_{B|A}$ as the quantum conditional state of $B$ given $A$.

Before drawing more analogies with classical probability theory, it is important to emphasize a subtlety involving conditional states. In the CSA framework, the conditional states can be divided into two different categories---expressing two different relations: those that connect regions that are \emph{acausally} related and those that connect regions that are \emph{causally} related. This distinction conveys the fact that two regions may be spatially separated or lie in the same spatial region but at different time instants, respectively. Consequently, conditional states linking acausal regions are called \emph{acausal conditional states}, and similarly, conditional states linking causally connected regions are called \emph{causal conditional states}. Although such relations lie at the very core of the CSA framework, we will not delve deeper into them here. For further clarification, we highly suggest Ref.~\cite{LS13}.

\subsection{Analogies Between Classical Probability Theory and the CSA}\label{SubSec.CausalAndAcausal}

We begin this subsection by establishing one of the pillars of CSA. Let $R$ be a random variable, and let the probability distribution over it be defined as $P(R)$ such that $\forall r \;\; P(R=r)\geq 0$ and  $\sum_r P(R=r) = 1$. Since classical probability obeys these conditions, we expect the analogous quantum operator to satisfy similar constraints. Accordingly, the corresponding operator to a classical probability distribution inside the CSA is a quantum state over a region $A$ satisfying\footnote{\label{footnote1} In this work, the set $\mathcal{D}(\mathcal{H}_X) =
\{\rho_{X} \in \mathcal{L}(\mathcal{H}_X)| \rho_{X} \geq 0,\text{Tr}(\rho_{X}) = 1\}$ represents the collection of density operators over $\mathcal{H}_X$, or simply the state space over $\mathcal{H}_{X}$. The notation $\rho_{X} \geq 0$ is usually to meant to say that this operator is positive semidefinite, that is, for all $\ket{\psi} \in \mathcal{H}_{X}$, $\bra{\psi}\rho\ket{\psi}\geq 0$ \cite{WatrousLectureNotes}.} $\rho_A \geq 0$ and $\patrace{\rho_A} = 1$, that is, $\rho_A \in \mathcal{D}(\mathcal{H}_{A})$. The requirement that the operator be positive semidefinite parallels the condition $\forall r \;\; P(R=r)\geq 0$, while the requirement $\patrace{\rho_A} = 1$ generalises the normalisation condition $\sum_r P(R=r) = 1$ \cite{LS13}.

As alluded to earlier, the CSA also provides several expressions that suggest a strong analogy between classical probability and quantum theory \cite{LS13}. Among those, one expression will be particularly useful in the second half of this paper: the \emph{quantum joint state}. That is, given a marginal state $\rho_A$ and a conditional state $\rho_{B|A}$, the joint state of the two regions $A$ and $B$ is a state $\rho_{AB} \in \mathcal{L}(\mathcal{H}_A \otimes \mathcal{H}_B)$ defined as,
\begin{equation}\label{Def.quantumjointstate}
    \varrho_{AB} := (\rho_{A}^{1/2} \otimes \mathds{1}_{B})\varrho_{B|A}(\rho_{A}^{1/2} \otimes \mathds{1}_{B}).
\end{equation}
Equation \eqref{Def.quantumjointstate} motivates the use of an auxiliary (binary) operation, usually defined as the \textit{star product}~\cite{LS13}.

\begin{definition}[$\star-$Product]\label{Def.starproduct}
Let $M,N$ be operators belonging respectively to the spaces $\mathcal{L}(\mathcal{H}_{A} \otimes \mathcal{H}_B)$ and $\mathcal{L}(\mathcal{H}_{A})$,\footnote{To be more concise, we will sometimes present the Hilbert space $\mathcal{L}(\mathcal{H}_{A} \otimes \mathcal{H}_B)$ as $\mathcal{L}(\mathcal{H}_{AB})$.} the star product between $M$ and $N$ is defined as:
\begin{equation}
\begin{aligned}
\star  :\mathcal{L}(\mathcal{H}_{AB}) \times \mathcal{L}(\mathcal{H}_{A}) &\rightarrow \mathcal{L}(\mathcal{H}_{AB}) \\
(M,N) &\mapsto M\star N,
\end{aligned}
\end{equation}
with the explicit form of the product,
\begin{equation}
    M\star N= (N^{1/2} \otimes \mathds{1}_{B}) M (N^{1/2} \otimes \mathds{1}_{B}).
\end{equation}
We also ask $N \geq 0 $ for its square root to be well-defined.
\end{definition}
Consequently, Eq.~\eqref{Def.quantumjointstate} can be concisely rewritten in the form:
\begin{equation}\label{Def.quantumjoitnstar}
    \varrho_{AB} = \varrho_{B|A} \star \rho_{A},
\end{equation}
which is analogous to the classical expression for a joint probability distribution \cite{probabilidade2023}---that is, given $R$ and $S$ random variables, their joint probability distribution is given as
\begin{equation}\label{ex.classicaljoint}
    P(R,S) = P(R|S)P(S).
\end{equation}


With the help of the Def.~\ref{Def.starproduct}, we can establish two more expressions that will also be specially helpful in the forthcoming calculations. The first shows how to obtain the marginal state $\rho_{B}$ from the joint state $\varrho_{AB}$:
\begin{equation}\label{Def.marginalstate}
    \rho_B := \text{Tr}_{A}(\varrho_{AB}).
\end{equation}
And the second is the \textit{quantum Bayesian inversion rule} for conditional states\footnote{In this work, we will tacitly admit that all operators have a proper inverse. When this is not the case, there is an easy workaround: if, say, $\rho_{B}$ were not invertible, we would restrict the respective equations involving $\rho_{B}^{-1}$ to supp$(\rho_{B})$. This is not inherently problematic in quantum theory, as conditionals $P(R|S)$ arising from $P(R,S)$ where $P(S=s)=0$ for some outcome $s$ are also present in classical probability.} which we define as follows:
\begin{equation}\label{Def.quantumbayes}
    \varrho_{A|B} := \varrho_{B|A} \star (\rho_{A}\rho_{B}^{-1}).
\end{equation}
As it is not our focus here, we will not delve deep into the construction of these expressions. The reader seeking a more comprehensive understanding can refer to Ref.~\cite{LS13}.

Finally, although Def.~\ref{Def.JamilIso} teach us how to connect a single quantum channel $\mathcal{N}_{B|A}$ with its respective conditional state $\varrho_{B|A}$, it is quite natural to ask how we can calculate the composition of two channels. In other words, given $\mathcal{N}_{B|A}$ and $\mathcal{N}_{C|B}$ we may want to determine what the Jamio\l kowski image of the composition $\mathcal{N}_{C|B} \circ \mathcal{N}_{B|A}$ is. The proposition below, whose proof can be found in Appendix \ref{App.ProofCompositionLaw}, addresses this question.

\begin{proposition}\label{Prop.chanelstatecomposition}
Let $\state{B}{A}$, $\state{C}{A}$, $\state{C}{B}$ be the Choi-Jamio\l kowski images of the CPTP maps $\channel{B}{A}$, $\channel{C}{A}$ and $\channel{C}{B}$ respectively. The composition 
    \begin{equation}
        \channel{C}{A}=\channel{C}{B} \circ \channel{B}{A},
        \label{Eq.PropMapsDiv}
    \end{equation}
holds true if and only if,
    \begin{equation}
        \state{C}{A} = \mbox{Tr}_{B}[(\mathds{1}_{A} \otimes \state{C}{B})^{T_B}(\state{B}{A} \otimes \mathds{1}_{C})].  
        \label{Eq.StatesDiv}
    \end{equation}
\label{Prop.DivChannelDivStates}
\end{proposition}

\subsection{Measurements}\label{SubSec.HybridStates}

Another important class of conditional states, especially in our case, is those associated with the physical processes of measurement. A quantum measurement $\mathcal{M}$ with classical outcomes $x$ in $Out(\mathcal{M})$ is represented~\cite{LS13,nielsen_chuang_2010,WatrousLectureNotes} by a POVM, a set $\{ E_{x}^{A}\}_{x \in Out(\mathcal{M})}$ of positive semidefinite operators acting on $\mathcal{H}_{A}$ such that 
\begin{align}
    \sum_{x \in Out(E)} E_{x}^{A} = \mathds{1}_{A}.
\end{align}

Since physical processes in quantum mechanics are represented by CPTP maps, the procedure of measuring a quantum state and obtaining an outcome should also have an associated quantum channel~\cite{WatrousLectureNotes}. If we restrict our attention to the classical outcomes and the probabilities of obtaining these outcomes, one way of representing a measurement is by a channel that takes a quantum state as input and outputs a classical state. In this sense, measurements can be represented as channels from a quantum region to a classical region. This class of physical process is represented within the CSA as a \emph{hybrid conditional state} \cite{LS13}. Such a class of states act on the composition of a classical region $X$ with a quantum region $A$. We clarify this discussion further in what follows. 

The Hilbert space $\mathcal{H}_X$ associated with a classical region $X$ comes with a preferred basis $\{\ket{x}\}_{x \in \text{Out(X)}}$ where all the density operators are represented by diagonal matrices on that basis \cite{LS13}---here, the set $\text{Out(X)}$ can be seen as a collection of labels for the outcomes of the measurement in question. Also, note that we do not require the quantum region $A$ to have a preferred basis. In this manner, the hybrid state associated with the measurement process, characterised by the CPTP channel $\varepsilon_{X|A}:\soplinear{A} \rightarrow \soplinear{X}$, is defined in the following way,
        \begin{equation}\label{Def.hybridmeasurment}
            \varrho_{X|A} := \sum_{x\in\text{Out}(X)}\ket{x}\bra{x} \otimes E_{x}^{A},
        \end{equation}
where the set of operators $\{E_{x}^{A}\}_{x \in Out(X)}$ is a POVM acting over $\mathcal{H}_A$.

One last comment. In this work, we have adopted a definition of hybrid conditional state associated with the measurement that is identical to the one provided in Ref.~\cite{LS13}, which will be particularly useful in the upcoming computations.   



\section{Schrödinger's cat as a remote measurement problem}\label{Sec.SCRMP}

    


\subsection{The Cat as a Remote Measurement Problem}\label{SubSec.Originalexperiment}

\begin{figure}
        \centering
        \includegraphics[width=0.7\linewidth]{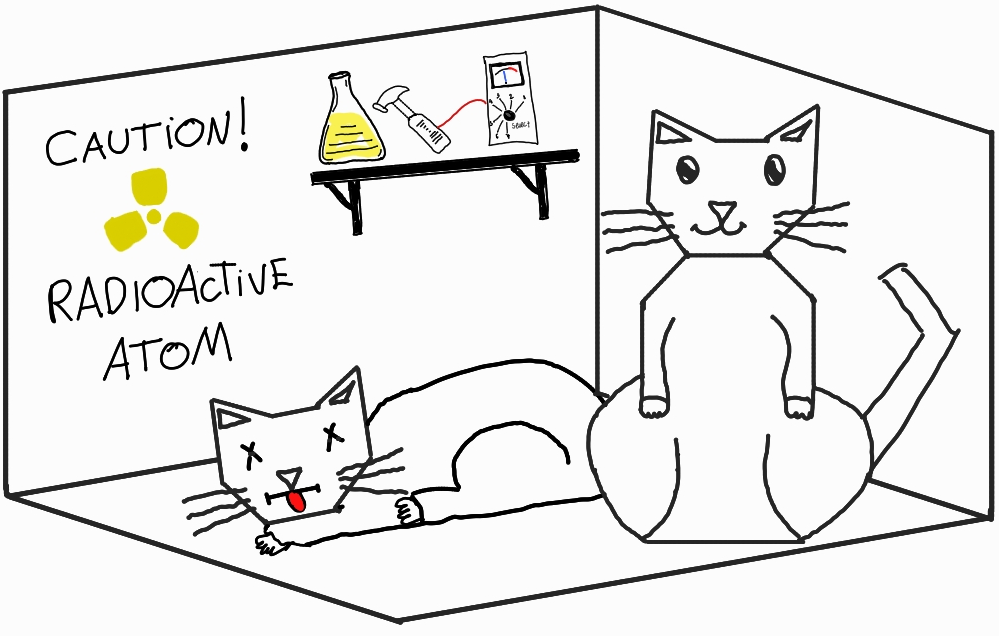}
        \caption{A not-so-fine work of art representing the original setup of Schrödinger's cat thought experiment. Taking a cross-section of the hermetically sealed box, we can see the representation of the ``diabolical device'' consisting of the Geiger counter, the poison, the hammer, and the radioactive atom. Also represented are the two possible states of the cat: alive and dead.}
        \label{fig_original_setup}
    \end{figure}

To recast Schrödinger's famous cat in a new light, we first review the original thought experiment. As originally proposed~\cite{trimmer1980present}, a cat is placed inside a hermetically sealed box together with a `diabolical device'. Such a contraption consists of a Geiger counter, a radioactive atom that may or may not decay during the course of the experiment, a hammer, and a bottle of poison, as represented in the Fig~\ref{fig_original_setup}. The cat is placed inside the box in such a way that it cannot interfere with the device. During the experiment, the Geiger counter may or may not register the decay of the atom. If it does, the counter triggers the hammer, which breaks the bottle and releases the poison, killing the cat. If it does not, the cat remains alive.

Since the box is hermetically sealed, the only way an observer can be sure of the cat's state is by directly intervening in the course of the experiment, for instance, by opening the box and eye-witnessing the actual states of affairs of the poor cat \cite{cohen,nielsen_chuang_2010}. With the box sealed, assuming that no disturbance occurs during the course of the experiment, the most accurate description an observer can provide is that the cat's state is in a superposition of `dead' and `alive', mathematically represented as:
    \begin{equation}
         \ket{\Psi} = \alpha \ket{\text{Dead}} + \beta \ket{\text{Alive}},
    \end{equation}
    with $|\alpha|^{2} + |\beta|^{2} = 1$.

\begin{figure}
        \centering
        \includegraphics[width=0.5\linewidth]{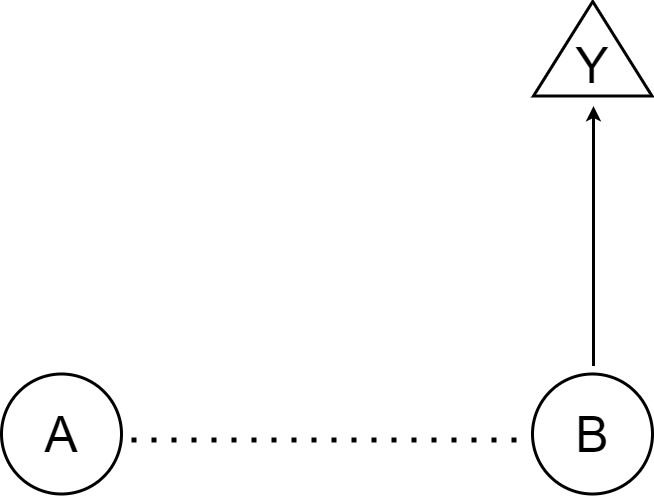}
        \caption{Diagrammatic representation of the recasting of Schrödinger's cat. Alice, located in region $A$, is in the presence of the cat, which was allowed to interact with the box for a limited time. The box is in region $B$, where Bob resides and can perform selected measurements on it. The dashed line between regions $A$ and $B$ indicates that they are spatially separated, while the continuous arrow from region $B$ to region $Y$ represents the measurement process that Bob can perform over the box and therefore, we have the same region at two different instants of time.}
        \label{fig.cat_box_classicla_simple}
    \end{figure}

Having brought the original experiment to light, we now move forward. We suppose, now, that the cat and the counter\footnote{Although we are referring to the cat-counter system, it is important to emphasise that the apparatus inside the box has more elements than just the counter. Essentially, upon detecting radioactive decay, the counter triggers a chain of events involving the hammer and the vial of poison. Therefore, we write the correlation as cat-counter, considering that the other elements of the system are embedded in this more compact notation.} are allowed to interact for a period of time $t$, during which their states become correlated. After this time $t > t_0$ (with $t_0$ being the initial time) has elapsed, the cat and the box are separated. The cat is sent to a room where Alice is located, while the hermetically sealed box is sent to another sufficiently distant room where Bob is. The cat and the box are now in two spatially separated regions. Moreover, given the particular experimental setup we are dealing with, the joint state of cat and counter is maximally entangled \cite{nielsen_chuang_2010,cohen}. It is due to the entanglement between the parts of the joint system that measurements realised on the counter state may be useful to infer the cat state---in a sort of spooky action at a distance, if one reifies the cat state. Such a system is represented in Fig.~\ref{fig.cat_box_classicla_simple}, where Alice stands in the $A$ region, together with the cat, and Bob stands in the $B$ region, together with the box.
    
Bob, who is in the presence of the box, has at his disposal a set of possible measurements he can perform on it. In particular, the measurements we consider are those in which Bob infers whether the radioactive atom has decayed or not. Once a specific measurement is chosen and realised, we may then ask: can Bob predict the state of the cat that is with Alice?

Although it might appear strange to a new reader that a cat and a box can reside in distinct regions while still sharing correlations, this type of scenario lies at the very core of quantum theory~\cite{einstein1935can,Bell1964,cohen}. This intrinsic correlation between the two subsystems enables measurements performed on one spatially separated part of the system to provide information about the other, a phenomenon usually referred to as \textit{remote measurement}. Although our proposed reframing of Schrödinger's cat experiment resembles the typical scenario of quantum steering, our reformulation differs slightly and is not an instance of the latter. See Ref.~\cite{UolaCostaNguyenGuehne2020} for an introduction to the subject and Ref.~\cite{LS13} for an alternative approach via quantum conditional states. 

The correlation between the cat and the box can be represented as an intrinsic relation between the cat’s state — namely, dead or alive — and the reading of the Geiger counter, whether it detects or does not detect the decay of the radioactive atom. Mathematically, we associate to the region where Alice lies with the cat (region A) a Hilbert space $\mathcal{H}_A$ with a basis $\beta_{\mathcal{H}_A} = \{\ket{D}, \ket{A}\}$, where $\ket{D}$ ($\ket{A}$) describes the state of the dead (alive) cat. To the region where Bob lies with the box (region B) we associate a Hilbert space $\mathcal{H}_B$ with a basis $\beta_{\mathcal{H}_B} = \{\ket{\uparrow}, \ket{\downarrow}\}$, where $\ket{\uparrow}$ ($\ket{\downarrow}$) corresponds to the reading of the counter, that is, if the atom has not decayed (has decayed). Thus, the composed region is represented by a Hilbert space $\mathcal{H}_{AB} = \mathcal{H}_A \otimes \mathcal{H}_B$ and the entangled state of the cat-counter system is then given as,
    \begin{equation}\label{Ex.CatChamVector}
        \ket{\psi}_{AB} = \frac{1}{\sqrt{2}}(\ket{\downarrow D} + \ket{\uparrow A}),
    \end{equation}
where $\ket{\psi}_{AB}$ coincides with the maximally entangled state $\ket{\Psi^{+}}$.
    
Our aim in this proposal is to determine the conditional state $\rho_{A|Y}$, which represents the state of the cat (located in region $A$) given the outcomes of the measurements performed on the box (stored in the classical region $Y$). Furthermore, we aim to determine the POVM associated with the possible measurements that Bob can perform on the box, and whether, by applying the POVM corresponding to the measurement he chooses, he can infer the state of the cat that lies in Alice’s presence.

\subsection{Mathematical Development}\label{SubSec.mathdev}
    Having now established the foundations of the analysis we wish to promote and having the state vector of the cat-box system established as in equation \eqref{Ex.CatChamVector}, we can utilise the quantum inference techniques provided by the CSA to search for the state $\rho_{A|Y}$.

    We begin by establishing the density operator associated with the state represented in the expression \eqref{Ex.CatChamVector},
    \begin{equation}\label{Ex.acausaljointinitial}
        \rho_{AB} = \kb{\psi}{\psi}_{AB}.
    \end{equation}
    
\begin{figure}
        \centering
        \includegraphics[width=0.6\linewidth]{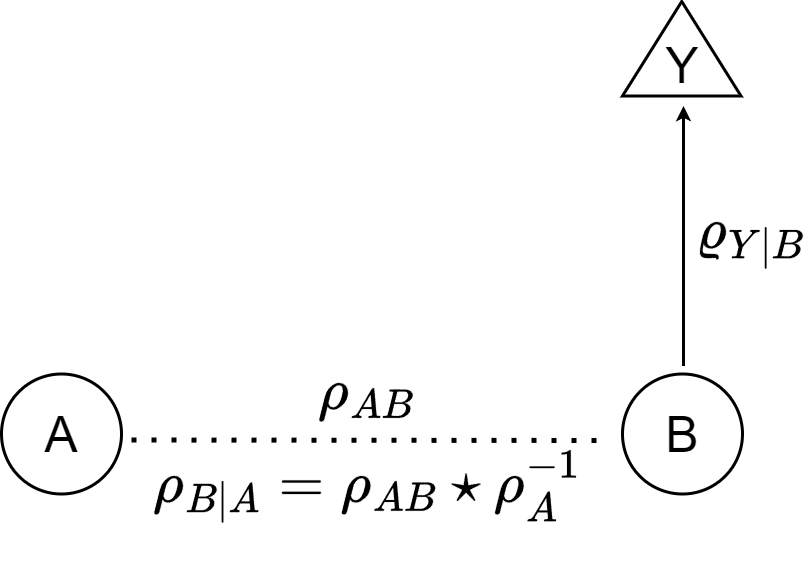}
        \caption{Diagrammatic representation of the recasting with the conditional states used to seek for our solution. In the lower part of the diagram we have the joint conditional state $\rho_{AB}$ between the regions of Alice and Bob and their associated conditional state $\rho_{B|A}$ obtained through the quantum rule for joint/conditional states. In the vertical branch of the diagram, we have the measurement process represented by the hybrid conditional state $\varrho_{Y|B}$.}
        \label{fig_reinterpretation_withCS}
    \end{figure}

Briefly returning to an earlier point. In Section~\ref{SubSec.CausalAndAcausal}, where we introduced an overview of CSA, we explained that conditional states have two different relations: those representing two spatially separated systems, and those representing a single system in the same region but at two different instants in time. Because of this, and to highlight this distinction, once the cat and the box are spatially separated, we choose to represent the conditional state associated with them with the Greek letter $\rho_{B|A}$ rather than $\varrho_{B|A}$, while the conditional hybrid state we kept the Greek letter $\varrho_{Y|B}$. This directly reflects the causal structure underlying CSA; however, we will not explore this topic further here, and refer the interested reader once again to Ref.~\cite{LS13} for additional details. Fig.~\ref{fig_reinterpretation_withCS} shows the diagram that represents the case we are investigating, together with the respective conditional and joint states.
    
Since our initial data is the state that represents the correlation between the regions, the density operator we obtain in the equation \eqref{Ex.acausaljointinitial} is precisely, within the CSA, our quantum joint state. Explicitly, the joint state of the regions $A$ and $B$ takes the form,
    \begin{equation}
        \begin{aligned}
            \rho_{AB} &= \frac{1}{2} (\ket{\downarrow D} + \ket{\uparrow A})(\bra{\downarrow D} + \bra{\uparrow A}) \\
            &= \frac{1}{2} (\kbt{\downarrow}{\downarrow}{D}{D} + \kbt{\downarrow}{\uparrow}{D}{A} + \\
            &+ \kbt{\uparrow}{\downarrow}{A}{D} + \kbt{\uparrow}{\uparrow}{A}{A}).
        \end{aligned}
    \end{equation}
   With the joint state in hand, we can use the quantum joint state rule (as defined in Eq.~\eqref{Def.quantumjoitnstar}) to obtain the conditional state of the region $B$ given the region $A$, namely $\rho_{B|A}$. That is, starting from the expression
    \begin{equation}
        \rho_{AB} = \rho_{B|A} \star \rho_{A},
    \end{equation}
    and taking the star product of $\rho_{A}^{-1}$ by the right on the both sides,
    \begin{align}
        &\rho_{AB} \star \rho_{A}^{-1} = \\ \nonumber
        &= (\rho_{A}^{-1/2} \otimes \mathds{1}_{B})(\rho_{A}^{1/2} \otimes \mathds{1}_{B}) \rho_{B|A}  (\rho_{A}^{1/2} \otimes \mathds{1}_{B})(\rho_{A}^{-1/2} \otimes \mathds{1}_{B}),
    \end{align}
    we arrive at
    \begin{equation}\label{eqrhoab}
        \rc{B}{A} = \rho_{AB} \star \rho_{A}^{-1}.
    \end{equation}
This expression tells us how to calculate the conditional state, given the joint state $\rho_{AB}$ and a marginal state $\rho_{A}$. 

To explicitly determine $\rho_{B|A}$, since we already have $\rho_{AB}$, we still need to obtain the marginal state of region $A$, namely $\rho_{A}$. To do so, we perform the marginalization of $\rho_{AB}$ as defined in expression~\eqref{Def.marginalstate}, and given that
    \begin{equation}
    \begin{aligned}
        Tr(\kb{\uparrow}{\uparrow}) &= Tr(\kb{\downarrow}{\downarrow}) = 1,   \\
        Tr(\kb{\uparrow}{\downarrow}) &= Tr(\kb{\downarrow}{\uparrow}) = 0,
    \end{aligned}
    \end{equation}
    we obtain that the marginal state is
    \begin{equation}\label{Ex.rho_A}
        \rho_{A} = \frac{1}{2} (\kb{D}{D} + \kb{A}{A}).
    \end{equation}
    Since $\rho_{A} \geq 0$, its inverse square root is given as
    \begin{equation}\label{rhoainv}
         \rho_{A}^{-1/2} =  \sqrt{2}(\kb{D}{D} + \kb{A}{A}).
    \end{equation}
    The marginal state $\rho_A$ and $\rho_{A}^{-1/2}$ can also be represented in their matrix form. Such representation is didactically useful to understand how the marginal state changes under the aforementioned operations. We have,
    \begin{equation}
        \begin{array}{cc}
            \rho_A = \begin{pmatrix}
                 \frac{1}{2} & 0 \\
                0 & \frac{1}{2}
                    \end{pmatrix}, &
        \rho_{A}^{-1/2} = \begin{pmatrix}
                 \sqrt{2} & 0 \\
                0 & \sqrt{2}
                        \end{pmatrix}.
        \end{array}
    \end{equation}
    
Substituting Eq.~\eqref{rhoainv} into Eq.~\eqref{eqrhoab}, we obtain the following conditional state:
    \begin{equation}
        \begin{aligned}\label{State.acausalBgivenA}
        \rc{B}{A} &= \kbt{\downarrow}{\downarrow}{D}{D} + \kbt{\downarrow}{\uparrow}{D}{A} + \\
         &+ \kbt{\uparrow}{\downarrow}{A}{D} + \kbt{\uparrow}{\uparrow}{A}{A}.
         \end{aligned}
    \end{equation}
    

From the characterisation provided by the expression \eqref{Def.hybridmeasurment} we know that the hybrid conditional state $\vrc{Y}{B}$ associated with the measurement that Bob will perform in the box, is of the form,
    \begin{equation}\label{State.hybridemeasurement}
        \rcc{Y}{B} = \sum_{y \in \{y_1,y_2\}} \kb{y}{y} \otimes E_{y}^{B},
    \end{equation}
    where $\{E^{B}_{y}\}$ is the POVM that Bob will work with. To the classical region $Y$, used to store the outcomes of the measurements performed on the box, we will associate a two-dimensional Hilbert space $\mathcal{H}_Y$ with a preferred basis $\beta_{\mathcal{H}_Y} = \{\ket{y_1},\ket{y_2} \}$. The choice of $\mathcal{H}_{Y}$ to be two-dimensional is naturally justified by the fact that we are considering that the counter can only register whether the atom has decayed or has not decayed. We will choose more appropriate labels and physical meaning in a minute. 
    
Now that we have the conditional state between regions $A$ and $B$, and a hybrid conditional state representing a dichotomic measurement performed by Bob, we can use Prop.~\ref{Prop.chanelstatecomposition} to compose these two states. Therefore, composing Eq.~\eqref{State.acausalBgivenA} and Eq.~\eqref{State.hybridemeasurement} we find the conditional state between the regions $A$ and $Y$, given by
    \begin{equation}\label{ex.rhoygivena}
        \rc{Y}{A} = Tr_B(\rcc{Y}{B} \rc{B}{A}).
    \end{equation}
   To show the full form of the expression \eqref{ex.rhoygivena}, we start by evaluating the term $\rcc{Y}{B} \rc{B}{A}$,
    \begin{equation}
        \begin{aligned}
            \rcc{Y}{B}\rc{B}{A} &= (\ttensor{\rcc{Y}{B}}{\mathds{1}_A})(\mathds{1}_Y \otimes \rc{B}{A}) \\
            &= \sum_{y \in \{y_1,y_2\}} (\kb{y}{y} \otimes E_{y}^{B} \otimes \mathds{1}_A)(\mathds{1}_Y \otimes \rc{B}{A}).
        \end{aligned}
    \end{equation}
    Taking the partial trace over the $B$ region we are led to,
    \begin{equation}
        \begin{aligned}
            \rc{Y}{A} = &\sum_{y \in \{y_1,y_2\}} \kb{y}{y} \otimes (\kbtre{D}{D}{\downarrow}{\downarrow} +\\ &+ \kbtre{D}{A}{\downarrow}{\uparrow} + \\ &+ \kbtre{A}{D}{\uparrow}{\downarrow} +\\&+ \kbtre{A}{A}{\uparrow}{\uparrow}).
        \end{aligned}
    \end{equation}
    Since we are aiming to establish the full form of the conditional state $\rho_{A|Y}$, and having set the conditional state $\rho_{Y|A}$, we can make use of the Bayesian inversion rule, as established in Eq.~\eqref{Def.quantumbayes}, to obtain the desired conditional state. That is,
    \begin{equation}\label{eqray}
        \rc{A}{Y} = \rcbayes{Y}{A}{A}{Y}.
    \end{equation}
 This quantum Bayesian inversion formula has a direct dependence on the marginal state $\rho_{Y}$ of the classical region. Such a marginal state can be obtained by applying belief propagation from region $B$ to region $Y$. Since we have access to both conditional and marginal states necessary to do this operation, that is, $\rcc{Y}{B}$ and $ \rho_B$, we obtain
    \begin{equation}
        \rho_Y = \text{Tr}_B(\rcc{Y}{B} \rho_B),
    \end{equation}
    where $\rho_B$ comes directly from the marginalization of the joint state $\rho_{AB}$, that is
    \begin{equation}
        \rho_B = \text{Tr}_A(\rho_{AB}) = \frac{1}{2}(\kb{\uparrow}{\uparrow} + \kb{\downarrow}{\downarrow}).
    \end{equation}
    Then, 
    \begin{equation} \label{state.rhoy}
        \begin{aligned} 
            \rho_Y &= \text{Tr}_B(\rcc{Y}{B}\rho_B) \\ 
                &=\text{Tr}_B \left\lbrace \frac{1}{2} (\sum_{y \in \{y_1,y_2\}} \kb{y}{y} \otimes E^{B}_{y})\lbrack \mathds{1}_Y \otimes (\kb{\uparrow}{\uparrow} + \kb{\downarrow}{\downarrow})\rbrack \right\rbrace\\ 
            &=\sum_{y \in \{y_1,y_2\}}\kb{y}{y} \text{Tr}(E_{y}^{B} \rho_B). 
        \end{aligned}
    \end{equation}
    The term $P_y = \text{Tr}(E_{y}^{B} \rho_B)$ that appears in the expression \eqref{state.rhoy} is the probability\footnote{As we will see later, once we define the POVM that Bob will use, both values of $P_Y$ are different from zero, thus avoiding possible complications in our solution.} of each one of the possibles outcomes of the measurement in the region $B$, and as we can see, it has a direct dependence on the POVM $\{E_{y}^{B}\}$ that Bob will choose. Therefore, the classical state $\rho_Y$ takes the form, 
    \begin{equation} 
        \rho_Y= \sum_{y \in \{y_1,y_2\}} P_y \kb{y}{y}, 
    \end{equation} 
    and the matrix representation of $\rho_Y$ (and $\rho_Y^{-1}$ ) are 
    \begin{equation} 
        \begin{array}{cc} 
            \rho_Y = \begin{pmatrix} 
                    P_{y_1} & 0 \\ 
                    0 & P_{y_2} 
            \end{pmatrix}, & 
        \rho_{Y}^{-1} = \begin{pmatrix} 
                P_{y_1}^{-1} & 0 \\ 
                0 & P_{y_2}^{-1} 
                        \end{pmatrix}. 
        \end{array}
    \end{equation}
    Finally, we obtain the desired conditional state of the region where the cat is located, given the region that stores the results of the measurements made on the box, that is
    \begin{equation} \label{Ex.finalformagivey}
        \begin{aligned} 
            \rc{A}{Y} &= \rcbayes{Y}{A}{A}{Y} \\ 
            &= \sum_{y \in \{y_1,y_2\}} \frac{P_{y}^{-1}}{2} \kb{y}{y} \otimes \lbrack \kbtre{D}{D}{\downarrow}{\downarrow} +\\&+ \kbtre{D}{A}{\downarrow}{\uparrow} + \\ 
            &+ \kbtre{A}{D}{\uparrow}{\downarrow} +\\&+ \kbtre{A}{A}{\uparrow}{\uparrow}\rbrack. 
            \end{aligned}
    \end{equation}
    
    To accurately infer Alice’s state, Bob must choose a POVM available to him, and it is only upon making this choice that he can accomplish his goal. Hence, we can determine the appropriate POVM that Bob might use in order to obtain his results. Since one of the simplest measurement Bob can perform on the box is to open it and verify whether the counter has registered one of the two possible outcomes—decayed or not—Bob therefore has at his disposal the following POVM: $\{\ket{\downarrow}\bra{\downarrow}, \ket{\uparrow}\bra{\uparrow}\}$. Where the POVM element $E_{Y=y_1}^{B} = \ket{\downarrow}\bra{\downarrow}$ is associated with the outcome ``decayed'', while the POVM element $E_{Y=y_2}^{B} = \ket{\uparrow}\bra{\uparrow}$ corresponds to the outcome ``not decayed''. And, as it should be,
    \begin{equation}
        \sum_{y \in \{y_1,y_2\}} E_{y}^{B} = \mathds{1}_B.
    \end{equation}

   Also, to check the self-consistency of the CSA, since we know the form of the marginal state $\rho_A$ given by expression~\eqref{Ex.rho_A} and we also have access to the POVM, we should expect that
    \begin{equation}\label{Ex.recoveringrho_A}
    \operatorname{Tr}_{Y}(\rho_{A|Y} \rho_Y) = \rho_{A}.
    \end{equation}
    To verify that equation~\eqref{Ex.recoveringrho_A} holds true, we begin by carefully analyzing the conditional state $\rho_{A|Y}$ with the POVM given above,
    \begin{align}\label{Ex.bigstatefullform} \nonumber 
    \rho_{A|Y} = \frac{P^{-1}_{y_1}}{2}\, \ket{y_1}\bra{y_1} \otimes &\Big[\operatorname{Tr}\!\left(E^{B}_{y_1}\ket{\downarrow}\bra{\downarrow}\right) \ket{D}\bra{D} \\\nonumber &+\operatorname{Tr}\!\left(E^{B}_{y_1}\ket{\downarrow}\bra{\uparrow}\right) \ket{D}\bra{A} \\ \nonumber 
    &+ \operatorname{Tr}\!\left(E^{B}_{y_1}\ket{\uparrow}\bra{\downarrow}\right) \ket{A}\bra{D} \\ \nonumber 
    &+\operatorname{Tr}\!\left(E^{B}_{y_1}\ket{\uparrow}\bra{\uparrow}\right) \ket{A}\bra{A} \Big] \\ \nonumber \quad + \frac{P^{-1}_{y_2}}{2}\, \ket{y_2}\bra{y_2} \otimes &\Big[\operatorname{Tr}\!\left(E^{B}_{y_2}\ket{\downarrow}\bra{\downarrow}\right) \ket{D}\bra{D} \\ \nonumber 
    &+\operatorname{Tr}\!\left(E^{B}_{y_2}\ket{\downarrow}\bra{\uparrow}\right) \ket{D}\bra{A} \\ \nonumber 
    &+\operatorname{Tr}\!\left(E^{B}_{y_2}\ket{\uparrow}\bra{\downarrow}\right) \ket{A}\bra{D} \\ 
    &+\operatorname{Tr}\!\left(E^{B}_{y_2}\ket{\uparrow}\bra{\uparrow}\right) \ket{A}\bra{A}\Big]. \end{align}

    By substituting the respective POVM elements, we obtain
    \begin{equation} \label{ex.reducedrhoaystateform}
     \rho_{A|Y} = \frac{P_{y_1}^{-1}}{2} \kb{y_1}{y_1} \otimes \kb{D}{D} + \frac{P_{y_2}^{-1}}{2} \kb{y_2}{y_2} \otimes \kb{A}{A}. 
    \end{equation}

    Finally, considering that $\rho_Y$ is given by Eq.~\eqref{state.rhoy}, we have
    \begin{align} \label{Ex.recoverrhoa} \nonumber
    \operatorname{Tr}_{Y}(\rho_{A|Y} \rho_Y) &= \operatorname{Tr}_{Y}\left\lbrack
    \frac{P_{y_1}^{-1} P_{y_1}}{2} \kb{y_1}{y_1} \otimes \kb{D}{D} \right. \\\nonumber
    &\left. + \frac{P_{y_2}^{-1} P_{y_2}}{2} \kb{y_2}{y_2} \otimes \kb{A}{A} 
    \right\rbrack \\
    &= \frac{1}{2}(\kb{A}{A} + \kb{D}{D}) = \rho_{A}.
    \end{align}

All in all, expression~\eqref{Ex.recoverrhoa} shows that, by propagating the belief from the classical region $Y$ to the quantum region $A$, where Alice is with the cat, we recover the correct state of the cat, namely $\rho_{A}$, as expected. This provides evidence that our conditional state $\rho_{A|Y}$ was correctly constructed.
    
    Now that we have $\rc{A}{Y}$ and the corresponding POVM, we can answer the question that permeates all this work. Suppose that Bob performs a measurement on the box and, say, the \textit{outcome} $y_1$ is the one that appears on the counter display after Bob has opened the box. How should the cat's state be updated in light of this new evidence? 

    With the measurement result in hand, Bob looks at the conditional state of region $A$ given region $Y$, as given by Eq.~ \eqref{ex.reducedrhoaystateform}, and selects the part of the conditional state that corresponds to the outcome he witnessed. In this particular case, given that the result was $Y=y_1$, Bob picks
    \begin{equation} \label{ex.actulizatedbobstate}
        \begin{aligned} 
        &\rc{A}{Y=y_1} = \frac{P_{y_1}^{-1}}{2}\kb{y_1}{y_1} \otimes \kb{D}{D}.
    \end{aligned}
    \end{equation}
    Also, since $P_Y = \text{Tr}(\rho_B E_{y}^{B})$, by the POVM established in the preceding paragraph, we set $P_{y_1} = P_{y_2} = 1/2$.
    
Bob, looking at the conditional state in Eq.~ \eqref{ex.actulizatedbobstate}, is able to infer that the state of the cat in Alice's presence is:
    \begin{align}\label{cat.dead}
        \rho_{A} = \ketbra{D}{D}.
    \end{align}

    Thus, we see that since Bob infers that the state of the box has been updated to ``decayed'', he can predict, by looking to the actualized conditional state in expression \eqref{ex.actulizatedbobstate} that the state of the cat in Alice's presence is ``dead''. If Bob had inferred that the box's state was ``not decayed'', he would then be able to predict that the cat's state is ``alive'', that is,
    \begin{equation}\label{cat.alive}
        \rho_A = \kb{A}{A}.
    \end{equation}

To conclude, note that the discussion above highlights the agent-centric nature of this framework. When Bob observes the result of the measurement, to predict the state of the cat that is with Alice, he necessarily has to look at the conditional state of the expression \eqref{ex.reducedrhoaystateform} and select one of the two possible parts of the conditional state. Only after selecting the part compatible with the result he has can Bob try to predict the cat's state in Alice's presence.

\section{Conclusion}\label{Sec.Conclusion}



In this work, we revisited and reframed Schrödinger's thought experiment using the framework provided by the conditional states approach. Initially, Schrödinger's "zombie cat" illustrated the seemingly counterintuitive concept of describing a single physical system as a superposition of two mutually exclusive states. However, because the conditional states formalism originated from an effort to interpret quantum theory as a subjective probabilistic assignment, we were able to recast the famous experiment as a remote measurement within an agent-centric framework. This perspective enabled us to look beyond the mystification surrounding the cat's well-being.
    

Within this approach, we demonstrated how the measurement process is constructed, and by applying the quantum Bayesian inversion rule, we established the conditional state that represents the cat’s state in Alice’s region, given the measurement outcomes obtained by Bob from his box. The argument is as follows: Bob and Alice share an entangled state representing the cat–counter system (see expression~\eqref{Ex.CatChamVector}). The conditional state associated with the cat–counter system in~\eqref{State.acausalBgivenA} is obtained through the quantum joint state rule ---see~\eqref{Def.quantumjointstate}. Bob, possessing the system whose hybrid state corresponds to the measurement he wishes to perform on the box, uses the state composition rule (Prop.~\ref{Prop.chanelstatecomposition}) to obtain the conditional state of the classical region that stores the outcomes of his measurements on the box, given Alice’s region. With $\rho_{Y|A}$ in hand, Bob then applies the quantum Bayesian inversion rule to obtain the desired conditional state $\rho_{A|Y}$ given in~\eqref{Ex.finalformagivey}. After performing the measurement on the box and acquiring new evidence, Bob updates his state to $\rho_{A|Y}$. Finally, based on the outcome from his measurement, Bob can determine whether the cat in Alice’s presence is either ``dead'' or ``alive'',  as given by expressions~\eqref{cat.dead} and~\eqref{cat.alive}.

Such an argument, as highlighted earlier, not only provides a useful construction but also emphasises the agent-dependent nature of this framework, as it clearly relies on Bob’s actions. It is Bob's beliefs about the cat's well-being that are superposed, and that will consequently suffer a posterior update upon his acquiring new information about the cat's actual state. We believe this is the probabilistic meaning that quantum theory should be about.

We hope that this contribution will serve not only as an introduction for new readers entering the field of quantum information and as a toy model of how to apply the CSA framework to a pivotal gedanken experiment, but also as a small contribution to the celebration of this Year of Quantum Technology.

\begin{acknowledgments}

We thank Carlos Humberto and Lucas Porto for the invaluable discussions. L L. Brugger thanks the support from the Fundação de Amparo à Pesquisa do Estado de Minas Gerais (FAPEMIG), process no. [15358] for his PhD scholarship and also thanks the Universidade Federal de Juiz de Fora for his master's scholarship. This work was supported by Grant 63209 from the John Templeton Foundation. The opinions expressed in this publication are those of the authors and do not necessarily reflect the views of the John Templeton Foundation. This work was supported by CNPq through two research grants from the Conhecimento Brasil Program (Lines 1 and 2). 

\end{acknowledgments}


%

\begin{widetext}

\appendix

\section{Proof of Proposition\ref{Prop.StateChoi}}\label{App.ProofIsomorphism}
\vspace{-0.7cm}
\begin{proof}
The ``if'' part. Suppose that $\rho$ verifies (a) and (b) of the proposition. We must prove that its isomorphic image $\mathcal{N} \circ \mbox{T}_{A}$ is completely positive and trace-preserving. 
Firstly, given $\sigma_{A} \in \mathcal{D}(\mathcal{H}_{A})$ note that:
{%
\setlength{\abovedisplayskip}{4pt}
\setlength{\abovedisplayshortskip}{4pt}
    \begin{align}
        \mbox{Tr}_{B}[\mathcal{N}_{B|A}(\sigma_{A}^{T})] &= \mbox{Tr}_{B}\mbox{Tr}_{A}[\rho_{B|A}(\sigma_{A}^{T} \otimes \mathds{1}_{B}  )] \nonumber \\
        &= \mbox{Tr}_{A}\mbox{Tr}_{B}[\rho_{B|A}( \sigma_{A}^{T} \otimes \mathds{1}_{B})] \nonumber \\
        &= \mbox{Tr}(\sigma_{A}^{T})= \mbox{Tr}(\sigma_{A}).
    \end{align}
}
So that $\mathcal{N}_{B|A} \circ \mbox{T}_{A}$ is trace-preserving. Now, it remains to prove that $\mathcal{N}_{B|A}\circ \mbox{T}_{A}$ is completely positive. Take $\sigma_{A} \in \mathcal{D}(\mathcal{H}_{A})$, then:
{%
\setlength{\abovedisplayskip}{4pt}
\setlength{\abovedisplayshortskip}{4pt}
\setlength{\belowdisplayskip}{4pt}
\setlength{\belowdisplayshortskip}{4pt}
    \begin{align}
    (\mathcal{N_{B|A}}\circ \mbox{T}_{A})(\sigma_{A}) &=  \mbox{Tr}_{A}\left[\rho_{B|A}( \sigma_{A}^{T} \otimes \mathds{1}_{B})\right] \\
    &= \mbox{Tr}_{A}\left[\sqrt{\rho_{B|A}^{T_{A}}}( \sigma_{A} \otimes \mathds{1}_{B})\sqrt{\rho_{B|A}^{T_{A}}}\right] \nonumber \\
        &=\sum_{a}(\bra{a} \otimes \mathds{1}_{B})\left(\sqrt{\rho_{B|A}^{T_{A}}}(\sigma_{A} \otimes \mathds{1}_{B})\sqrt{\rho_{B|A}^{T_{A}}}\right) (\ket{a} \otimes \mathds{1}_{B}) \nonumber \\
        &=\sum_{a}\left[(\bra{a} \otimes \mathds{1}_{B})\sqrt{\rho_{B|A}^{T_{A}}}\right](\sigma_{A} \otimes \mathds{1}_{B})\left[\sqrt{\rho_{B|A}^{T_{A}}}(\ket{a} \otimes \mathds{1}_{B})\right].
    \end{align}
}
Here, $\{\ket{a}\}_{a=1}^{d_a}$ is an orthonormal basis of $\mathcal{H}_{A}$. Defining, $K_{a}:=[(\bra{a} \otimes \mathds{1}_{B})\sqrt{\rho_{B|A}^{T_{A}}}]$ for all $a$, we can rewrite $\mathcal{N}_{B|A}(\sigma_{A})$ as
{%
\setlength{\abovedisplayskip}{4pt}
\setlength{\abovedisplayshortskip}{4pt}
\setlength{\belowdisplayskip}{4pt}
\setlength{\belowdisplayshortskip}{4pt}
\begin{equation}\label{Ex.ka}
    \mathcal{N}_{B|A}(\sigma_{A})=\sum_{a}K_{a}(\sigma_{A} \otimes \mathds{1}_{B})K_{a}^{\ast},
\end{equation}
}
where operators of the form presented in expression \eqref{Ex.ka} are CP. This concludes the first half of the proof. 

The ``only if'' part. Assume that $\mathcal{N}_{B|A} \circ \mbox{T}_{A}$ is completely positive and trace-preserving. For this part, we use the inversion expressed in~\eqref{Eq.DefCJIsomorphism}. As a matter of fact,
{%
\setlength{\abovedisplayskip}{4pt}
\setlength{\abovedisplayshortskip}{4pt}
\setlength{\belowdisplayskip}{4pt}
\setlength{\belowdisplayshortskip}{4pt}
\begin{align}
    \mbox{Tr}_{B}(\rho_{B|A}) &= \mbox{Tr}_{B}\left[\sum_{i,j} \ket{i}\bra{j} \otimes \mathcal{N}_{B|A}(\ketbra{j}{i})\right]  \nonumber \\
    &= \sum_{i,j} \ket{i}\bra{j} \otimes \mbox{Tr}\left[\mathcal{N}_{B|A}\circ \mbox{T}_{A}(\ketbra{i}{j})\right]  \\
    &= \sum_{i,j} \ket{i}\bra{j}[\mbox{Tr}\ketbra{j}{i})] \nonumber \\
    &=\sum_{i,j}\ketbra{i}{j}\delta_{i,j}= \sum_{i} \ket{i}\bra{i}= \mathds{1}_{A}. \nonumber
\end{align}
}
\noindent 
and that is saying that Tr$_{B}(\rho_{B|A})=\mathds{1}_{A}$. Now, we have to prove item (a). That is done by noticing that $\rho_{B|A}^{T_{A}} = \mbox{id} \otimes (\mathcal{N}_{B|A}\circ \mbox{T}_{A}
)(\Phi^{+})$, where $\Phi^{+}:=\sum_{i,j}\ketbra{i}{j} \otimes \ketbra{i}{j}$ is the usual (non-normalised) Bell-state. As the composition $\mathcal{N}_{B|A} \circ \mbox{T}_{A}$ is completely positive, it implies that $\rho_{B|A} \geq 0$.
\end{proof}

\section{Proof of Proposition \ref{Prop.DivChannelDivStates}}\label{App.ProofCompositionLaw}
\vspace{-0.3cm}
\begin{proof}
The "if" part. Assume that expression~\eqref{Eq.PropMapsDiv} holds true. In this case:
{%
\setlength{\abovedisplayskip}{4pt}
\setlength{\abovedisplayshortskip}{4pt}
\setlength{\belowdisplayskip}{4pt}
\setlength{\belowdisplayshortskip}{4pt}
\begin{equation}
    \channel{2}{0}=\channel{2}{1} \circ \channel{1}{0}: \mathcal{L}(\mathcal{H}_{0}) \rightarrow \mathcal{L}(\mathcal{H}_{2})
\end{equation}
}
is a completely positive trace-preserving map arising from the composition of the other two CPTP maps. For the sake of comprehension, denote $\rho_{i|j}$ as $\mathcal{J}(\mathcal{N}_{i|j})$. Then,
\begin{align}
    \varrho_{2|0} \circ \mbox{T}_{\mathcal{H}_{0}} &=  \rho_{2|0} = \mathcal{J}(\mathcal{N}_{2|0}) =   \mathcal{J}(\mathcal{N}_{2|1} \circ \mathcal{N}_{1|0}) \nonumber \\
    &=\left[\mbox{id} \otimes (\mathcal{N}_{2|1} \circ \mathcal{N}_{1|0})\right]\left(\sum_{i,j}\ketbra{i}{j} \otimes \ketbra{j}{i}\right) \nonumber \\
    &= (\mbox{id} \otimes \mathcal{N}_{2|1})(\mbox{id} \otimes \mathcal{N}_{1|0})\left(\sum_{i,j}\ketbra{i}{j} \otimes \ketbra{j}{i}\right)  \\
    &= (\mbox{id} \otimes \mathcal{N}_{2|1})(\rho_{1|0}) \nonumber \\
    & = \mbox{Tr}_{\mathcal{H}_{1}}\left[ (\mathds{1}_{\mathcal{H}_{0}} \otimes \rho_{2|1}) ( \rho_{1|0} \otimes \mathds{1}_{\mathcal{H}_{2}}) \right], \nonumber \\
    & = \mbox{Tr}_{\mathcal{H}_{1}} \left[ (\mathds{1}_{\mathcal{H}_{0}} \otimes \rho_{2|1}) ( \varrho_{1|0} \otimes \mathds{1}_{\mathcal{H}_{2}})^{T_{\mathcal{H}_{0}}} \right] \nonumber \\
    & = \left\lbrace\mbox{Tr}_{\mathcal{H}_{1}} \left[ (\mathds{1}_{\mathcal{H}_{0}} \otimes \rho_{2|1}) ( \varrho_{1|0} \otimes \mathds{1}_{\mathcal{H}_{2}}) \right]\right\rbrace^{T_{\mathcal{H}_{0}}}.
\end{align}
In conclusion, %
\begin{equation}
    \varrho_{2|0}=\mbox{Tr}_{\mathcal{H}_{1}} \left[ (\mathds{1}_{\mathcal{H}_{0}} \otimes \varrho_{2|1})^{T_{\mathcal{H}_{1}}} ( \varrho_{1|0} \otimes \mathds{1}_{\mathcal{H}_{2}}) \right].
\end{equation}

The "only if" direction follows in complete analogy, and we will not write it down here. 

\end{proof}

\end{widetext}
\end{document}